# Influence of Critical Current Distribution on Operation, Quench Detection and Protection of HTS Pancake Coils

M. Wozniak, E. Schnaubelt, S. Atalay, B. Bordini,
J. Dular, T. Mulder, E. Ravaioli and A. Verweij

*Abstract*— High-temperature superconductor (HTS) coated conductors (CC) are often wound into pancake coils with electrical insulation in-between the turns. The copper terminals are used for current injection and conduction cooling. An inherent variation of the critical current along the CC length results from its manufacturing process. This variation causes non-uniform heat generation, particularly when the coil is operated at a high fraction of the nominal critical current or when large critical current defects are present. The temperature distribution resulting from the balance between cooling and heating, in combination with the magnetic field and critical current distributions, determines whether a thermal runaway occurs. Accurately predicting the level of critical current defects that can be tolerated during conduction-cooled operation is difficult and requires a three-dimensional (3D) coupled electromagnetic and thermal simulation.

This paper presents the results of simulations that are performed with the open-source Finite Element Quench Simulator (FiQuS) tool developed at CERN as part of the STEAM framework. The 3D coupled magnetodynamic-thermal simulations are based on the H− ϕ formulation and use thin shell approximations, a CC homogenization and conduction-cooling. The critical current can be varied along the CC length.

The effect of a single defect specified as a reduction of the critical current along the CC length is investigated in terms of the coil's ability to reach and maintain the operating conditions. The critical current and length of the defect that results in a thermal runaway are analyzed in terms of defect location in the coil. In addition, a classical one-dimensional scenario with a quench heater is studied. Both the local defect and the heater cases are compared in terms of the voltage signal available for quench detection. These cases result in very different requirements for quench detection, and their implications are discussed.

## I. INTRODUCTION

RECENT advancements in high-temperature superconductor (HTS) coated conductors (CC) have improved performance, reduced costs, and increased the availability of long-length conductors and manufacturing capacity. These developments make HTS CC an increasingly viable alternative to low-temperature superconductors (LTS) [1, 2]. In particle accelerators, moving away from superfluid or bath cooling for the majority of magnets could significantly reduce energy consumption and improve sustainability, particularly in machines like the Future Circular Collider [3]. However, helium gas or conduction-cooled HTS magnets require distinct design approaches to ensure stable operation and effective quench detection and protection — key topics explored in this paper.

While accelerator-class HTS magnets are not yet available for comprehensive studies, certain simpler configurations, such as single-CC pancake coils, can be studied in detail and offer valuable insights into how conduction-cooled HTS magnets behave. No-insulation (NI) coils [4] have been explored, but concerns about their increased power dissipation during ramping [5, 6] and reduced field quality [7, 8] make insulated coils a more attractive option.

A challenge with HTS CCs is the variation in critical current, $I_c$, along their length, typically due to the manufacturing process. Local decreases in $I_c$, referred to as defects, are a particular concern and are routinely measured as part of quality control [9, 10]. However, determining which defects are acceptable for stable operation is complex and not straightforward. This complexity often encourages strict quality assurance specifications regarding $I_c$, which may reduce manufacturing yields and increase costs. Few studies have explored how to quantify acceptable $I_c$ defects for conduction-cooled coils [11], highlighting an important gap in the research.

On the other hand, HTS CC studies often rely on one-dimensional (1D) models to simulate quench behavior, particularly focusing on minimum quench energy (MQE). In these models, quenching is typically assumed to begin at a single localized point, with quench propagation studied along a single axis. Although the MQE is often found to be relatively large, in the range of joules, point-like heat deposition is still considered a valid approach for quench initiation. However, quench dynamics differ significantly depending on the reason for the quench—whether it is dynamically initiated by a heater or statically triggered by a local critical current reduction.

M. Wozniak, E. Schnaubelt, S. Atalay, B. Bordini, J. Dular, T. Mulder, E. Ravaioli and A. Verweij are with CERN, Meyrin, Switzerland (e-mail: mariusz.wozniak@cern.ch).

Recent advancements, such as the Finite Element Quench Simulator (FiQuS) [12], developed as part of CERN's STEAM [13] framework, enable the use of three-dimensional (3D) models. These 3D simulations consider the full geometry and compute a resistive voltage for quench detection under various thermal conditions, providing deeper insights into thermal runaway scenarios. FiQuS can also simulate local $I_c$ degradation and its effects in a coupled magnetodynamic-thermal environment.

In this paper, acceptable $I_c$ defect lengths, positions and degradation levels are presented for stable operation in conduction-cooled coil geometry, and factors influencing the distribution of acceptable $I_c$ defects are explored. Finally, the efficiency of resistive voltage-based quench detection is presented, and the traditional 1D heater-induced quench is compared with the more advanced 3D defect-induced quench. While quench detection is closely tied to protection methods, this paper will not delve into the details of quench protection.

## II. PANCAKE COIL

A single, 4 mm wide (Re)BCO CC with a copper stabilizer is considered, and its key properties are listed in Table I. The CC has 35 µm Kapton insulation applied to both sides. Material properties of Kapton used in the simulations were [14, 15].

TABLE I
KEY PARAMETERS OF THE (RE)BCO CC

| Parameter | Value | Unit |
|---|---|---|
| $I_c$ (15 K, 15 T) | 700 | A |
| n-value = const. | 30 | - |
| Width | 4 | mm |
| (Re)BCO thickness | 2.4 | µm |
| Hastelloy® thickness | 40 | µm |
| Copper thickness | 60 | µm |
| Silver thickness | 2 | µm |
| Copper / Silver RRR | 100 / 20 | - |
| Insulation material | Kapton | - |
| Insulation thickness | 35 | µm |
| Winding cell thickness | 174.4 | µm |

As listed in Table II, a single, 20-turn pancake coil with a winding inner and outer diameter (I.D. and O.D.) and copper terminals (Fig. 1) is studied. A background magnetic field, $B_b$ of 15 T, is applied in the axial direction. At an operating current of 500 A, the coil generates $B_c$, a central field $B_{cc}$ of 0.5 T, and a peak field $B_{cp}$ of 1.3 T. At the nominal temperature $T_n$ of 15 K and the coil maximum field $B_{max}=B_b+B_{cp}=16.3$ T, the operating current, $I_{op}$ of 500 A, corresponds to 76% of $I_c(T_{op}, B_{max})$ of the CC.

## III. SIMULATION MODEL

### A. Finite Element tool

The open-source FiQuS tool is coded in Python [16] and, with the help of Gmsh [17, 18] and GetDP [19, 20], performs geometry generation, meshing, solving, and post-processing.

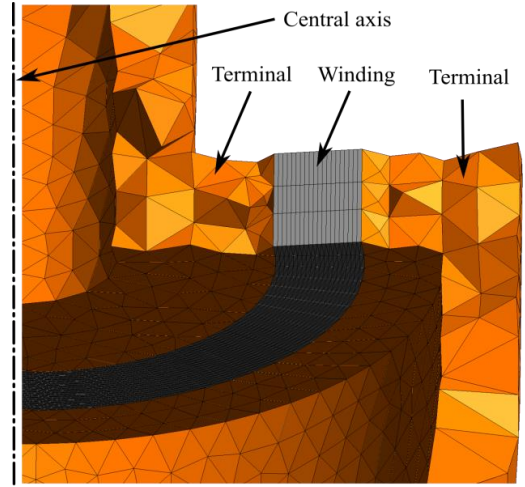

Fig. 1 Cross-section of the pancake coil mesh, with each CC meshed as 3 elements in the axial and one in the radial direction.

TABLE II
KEY PARAMETERS OF THE PANCAKE COIL

| Parameter | Value | Unit |
|---|---|---|
| Number of turns | 20 | - |
| Winding I.D. / O.D. | 20 / 26.98 | mm |
| Terminals thickness/length | 3 / 30 | mm |
| Terminals material/RRR | Cu / 100 | - |
| Conductor length | 1.487 | m |
| Operating current, $I_{op}$ | 500 | A |
| Fraction of $I_{op} / I_c$ ($T_n, B_{max}$) | 76 | % |
| Winding cell current density at $I_{op}$ | 717 | A/mm² |
| Nominal temperature, $T_n$ | 15 | K |
| Axial background field, $B_b$ | 15 | T |
| Coil center/max field | 15.5/16.3 | T |

The Pancake3D module of FiQuS version 2024.10.3 [21] was used with insulated coil capabilities as described in [22]. The insulation layer between the turns was simulated using the magneto-thermal thin shell approximation (TSA) [23, 24, 25]. The thermal and magnetodynamic contributions to the model behavior were adjusted to study their impact independently. By default, the magnetic field and temperature distribution are calculated in 3D, referred to as 3D cases in the following. To simulate classical 1D thermal heat diffusion in the winding, the heat flow between coil turns (i.e., across the thermal TSA) was turned off for the cases referred to as 1D in the following. Note that the heat flow modeled in the 1D case is two-dimensional since the axial heat flow is considered. The temperature gradient in the axial direction is negligible, and the thermal simulation is essentially 1D. The term 1D is used instead of 2D to highlight that this case simulates a classical 1D quench propagation. $I_c$ was considered as either $I_c(B_b, T)$ or $I_c(B_b+B_c, T)$ for the cases with background field only or with total magnetic field.

*B. $J_c(B, T)$ function*

The critical current density was defined as [26]:

$$J_c(B,T) = \frac{I_{c,0}(T)}{A_{sc}}\left(1 + \frac{B}{B_{0,0}(T)}\right)^{-\alpha}\left(1 - \frac{B}{B_{irr,0}(T)}\right)^{-q}, \quad (1)$$

with the temperature-dependent parameters given by:

$$I_{c,0}(T) = I_{c,0,0}\left(1 - \frac{T}{T_{c,0}}\right)^{\gamma}, \quad (2)$$

$$B_{0,0}(T) = B_{0,0,0}\left(1 - \frac{T}{T_{c,0,0}}\right), \quad (3)$$

$$B_{irr,0}(T) = B_{irr,0,0}\left(1 - \frac{T}{T_{c,0,0}}\right)^{\beta}, \quad (4)$$

with $A_{sc}$ the cross-section of the (Re)BCO layer and α, β, γ and $q$ fitting constants.

The $J_c(B, T)$ fit was multiplied by $A_{sc}$ to obtain $I_c(B, T)$, and a fit was performed at 0 deg (field perpendicular to the wide surface of the CC) to data of the CC manufactured by the Faraday Factory and available at [27].

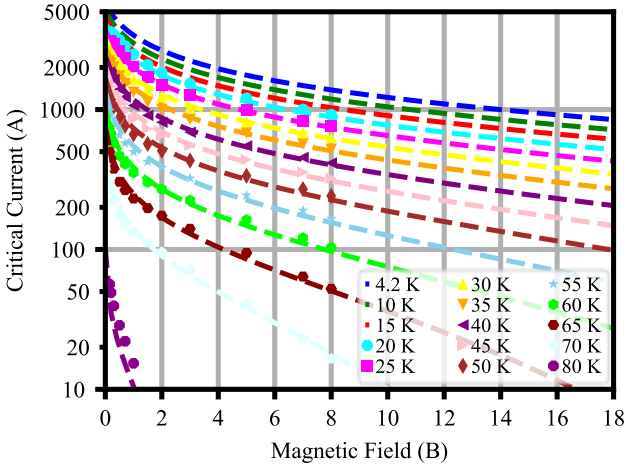

Fig. 2 Critical current as a function of magnetic field parameterized by the temperature. Measured data points are represented by markers, and the dashed lines represent the $I_c(B, T)$ fit of eq. (1) with parameters given in Table III.

TABLE III
$J_C(B, T)$ FIT PARAMETERS OF EQUATIONS (1)-(4)

| Parameter | Value | Unit |
|---|---|---|
| $I_{c,0,0}$ | 6098.7 | A |
| $B_{0,0,0}$ / $B_{irr,0,0}$ | 0.5769 / 200 | T |
| $T_{c,0,0}$ | 85 | K |
| $A_{sc}$ | 9600 | μm² |
| α / β / γ / q | 0.48175 / 1.4 / 1.45 / 2.0 | - |

The data and the fit are plotted in Fig. 2, and the fit parameters are listed in Table III. The $J_c(B, T)$ function is open-source [28] and was compiled with CERNGetDP [29] (version 2024.8.2) and used in this study. The function can be used in other software, such as COMSOL Multiphysics, Python or MATLAB [30].

*C. Powering and cooling setup*

The coil was simulated by the schematic shown in Fig. 3. The current source was switched off when the coil peak temperature $T_{max}$ reached 300 K. The coil was conduction cooled via its terminals, with their ends fixed at the nominal temperature $T_n$ of 15 K. The heat dissipation in the terminals was disregarded, corresponding to a superconducting shunt for the current path. The temperature gradient in the terminals between the coil and the cooling source is taken into account.

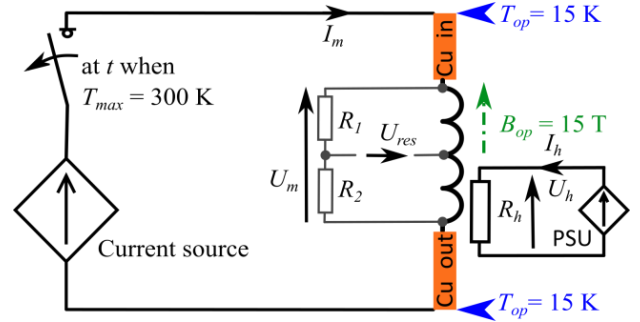

Fig. 3 Simplified schematic of the magnet electrical and thermal circuits.

*D. Resistive voltage computation*

The resistive component of the coil voltage is used for quench detection, achieved by using a central voltage tap (Fig. 3) with ideal compensation of the inductive voltage. To compute the resistive voltage $U_{res}$ of the HTS CC winding as a post-processing quantity, a solid conductor winding function $\chi = -\nabla v$ with 3D support in the whole coil was computed [31]. The electric scalar potential $v$ was calculated in each time step by solving a unitary stationary current flow problem following [32] using the conductivity values for the computed temperature and magnetic field distribution at that time step. From $\chi$, the DC conductance $G$ was computed. Finally, $U_{res}$ was obtained by multiplying $G$ with the total voltage $U_m$ and the potential difference between the beginning and the end of the winding, i.e., the difference of values of $v$ at those locations.

*E. Defect definition and quench initiation*

A defect is a region of the CC with a reduced critical current due to a local reduction of the (Re)BCO thickness or width. This means that the critical current of the defect, $I_{c,d}(B, T)$, is changing with field and temperature. A single defect is introduced by locally changing the parameter $I_{c,0,0}$ in eq. 2, such that the local $I_c(B_b, T_n)$ or $I_c(B_b+B_c, T)$ matches $I_{c,d}$.

The defect length, $l_d$, and its central location in terms of coil turn number $n_d$ need to be specified for a full defect definition (Fig. 4). The turn number $n_d$ is equivalent to specifying a location along the conductor length, $x$, although the relation between $n_d$ and $x$ is nonlinear as shown in Fig. 4. The first ($n_d \in [0,1]$) and last turn ($n_d \in [19,20]$) are soldered to the copper terminals. Defects in these locations and up to 0.5 turns away are not considered; hence $n_d$ is in the range from 1.5 to 18.5. A scenario without defect is also studied where a virtual heater initiates a quench. In this case, a volumetric power density is applied in the CC over a heater length $l_h = l_d$ and at location $n_h = n_d$.

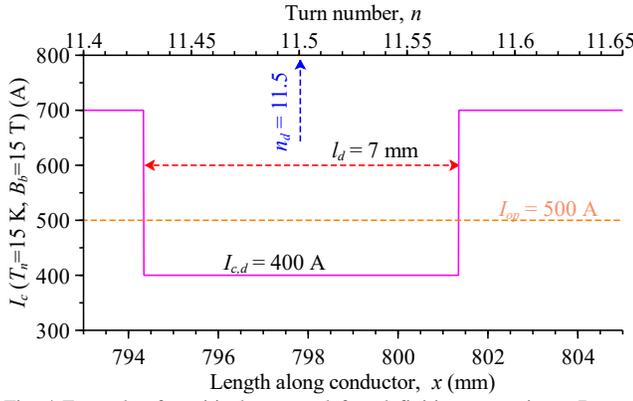

Fig. 4 Example of a critical current defect definition, assuming a 7 mm long defect at a location of turn 11.5 and with a critical current of 400 A.

## IV. RESULTS AND DISCUSSION

Section IV.A introduces the concept of the minimum stable defect critical current $I_{c,d,m}$ for fixed defect length. Next, the dependence of $I_{c,d,m}$ on $n_d$ is examined for both 1D and 3D cases (Section IV.B). Following that, the impact of a larger $l_d$ and $B_c$ on $I_{c,d,m}$ is discussed (Section IV.C). Finally, the coil resistive voltage for quench detection is presented when $I_{c,d}$ is just below $I_{c,d,m}$. This is then compared with quench initiation using a heater, and the differences are analyzed (section IV.D).

### A. Minimum stable defect critical current

The minimum stable defect critical current, $I_{c,d,m}$, is defined as the value of $I_{c,d}$, which allows the coil to reach its $I_{op}$, and operate without experiencing a thermal runaway. The $I_{c,d,m}$ is determined for coil current ramp rates that do not induce substantial AC losses, which could otherwise lead to thermal runaway during the ramp.

Fig. 5 shows the peak coil temperature for $l_d = 2$ mm at turn $n_d = 11.5$ and with $I_{c,d}$ of 476 and 477 A. The lower value causes a thermal runaway, whereas the coil peak temperature and total Joule power are stable for the higher $I_{c,d}$ of 477 A, which therefore corresponds to the $I_{c,d,m}$ for the $l_d$ and $n_d$ considered.

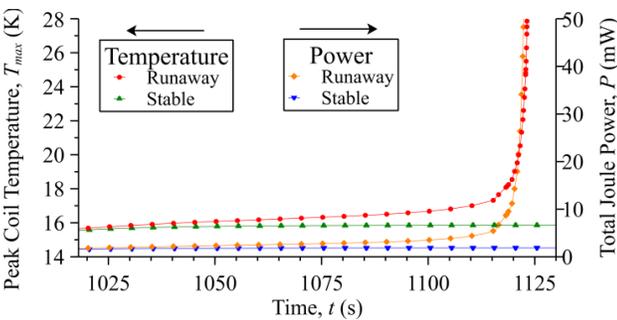

Fig. 5 Peak temperature and total Joule power in the coil. The stable (runaway) $I_{c,d}$ is 477 (476) A. The time range close to the runaway is shown.

### B. Defect location and 1D or 3D case

A study of the dependence of $I_{c,d,m}$ on $n_d$ with $l_d = 2$ mm was performed for 1D and 3D cases (Fig. 6). In the 1D case, the $I_{c,d,m}$ ranges from 349 to 477 A. The dependence is largely due to the heat conduction path length from the defect to the (cooling) terminals. The shortest length is for $n_d=1.5$, which results in the lowest $I_{c,d,m}$. The 3D case allows for heat diffusion between the turns, which changes the $I_{c,d,m}$ dramatically. For $n_d$ in the range 1.5-6.5 and 15.5-18.5 turns, the $I_{c,d,m}$ is zero, and for turns in the middle of the coil, it is between 50 and 58 A.

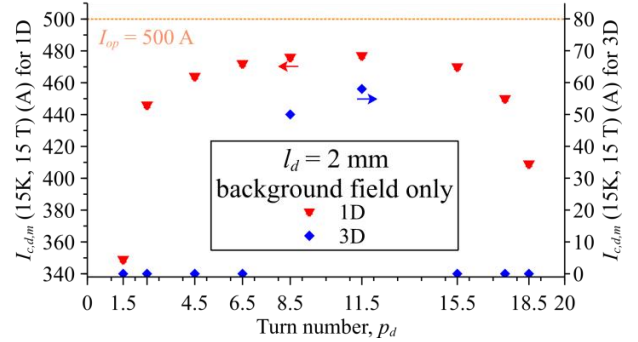

Fig. 6 Minimum stable defect critical current $I_{c,d,m}$(15 K, 15 T) calculated for 1D or 3D cases. Both are with the background magnetic field $B_b$ only.

### C. Longer defect in 3D and coil magnetic field impact

Increasing $l_d$ would increase $I_{c,d,m}$, and for $l_d>7$ mm $I_{c,d,m}$ is larger than 0 A for any considered location in the coil. As an example, $I_{c,d,m}$ is plotted in

Fig. 7 for the 3D case for $l_d$ of 7 mm. The most stable position, with the lowest $I_{c,d,m}$ is now on the outside of the coil ($n_d=18.5$). In this area, heat conduction across the insulation dominates because the turn radius is larger, hence the insulation cross-section for the heat conduction.

The coil peak field $B_{cp}$ is 1.3 T at $I_{op}$. When $B_c$ is included together with the $B_b$, it impacts $I_{c,d,m}$, which is increased at the inner turns and decreased at the outer turns (

Fig. 7).

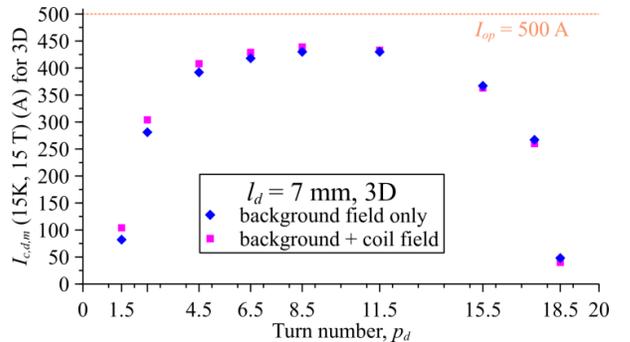

Fig. 7 Minimum stable defect critical current $I_{c,d,m}$(15 K, 15 T) calculated for case heat diffusion in 3D and with background field $B_b$ only or with background field and coil field $B_b+B_c$.

### D. Quench detection and thermal runaway

To cause a thermal runaway, an $I_{c,d}$ of 2 A below the $I_{c,d,m}$ for $l_d=2$ mm and $n_d=11.5$ was chosen. Of key interest is the time from the resistive voltage crossing the 0.1 V quench detection threshold until the time when the thermal runaway is well advanced, which is taken when the coil peak temperature exceeds 200 K.

Fig. 8 shows the last 200 s before a thermal runaway during a current ramp to 500 A in 2400 s (40 min) for the 3D case.

The voltage during the ramp steadily increased and reached 0.1 V about 36.9 s before the thermal runaway.

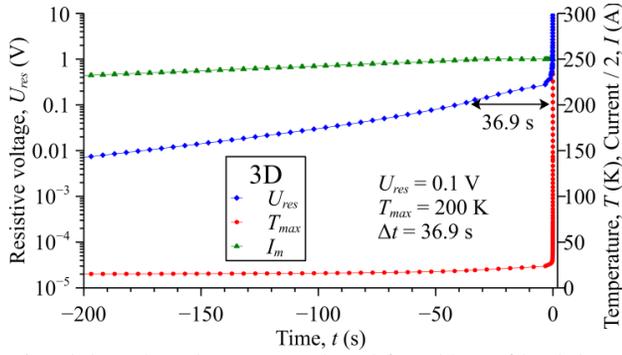

Fig. 8 Resistive voltage due to a 7-mm long defect with $I_{c,d}$ of 2 A below $I_{c,d,m}$ and location of turn 11.5 with 3D heat diffusion and background magnetic field only. The time axis was adjusted to be zero when $T_{max}$ reached 300 K.

The typically studied 1D case with heater pulse quench initiation was used to simulate a thermal runaway in Fig. 9 caused by a heater pulse just above the MQE of the CC in the 1D case. As in the 3D case with a local defect, the coil current was $I_{op}$ when the heater power was applied, and a runaway occurred. The total resistive voltage crosses 0.1 V and is followed by a thermal runaway only after 20 ms, showing that there is very little time to protect the coil.

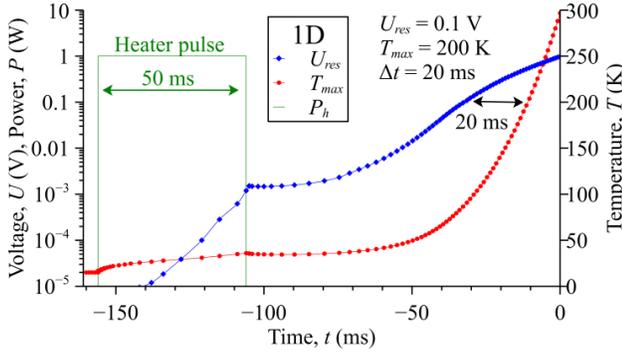

Fig. 9 Resistive voltage due to a heater-induced quench with a heater length of 7 mm at the location of turn 11.5 with 1D heat diffusion and background magnetic field only. The time axis was adjusted to be zero when $T_{max}$ reached 300 K.

The 'quasi-static' quench initiation due to a defect with 3D heat diffusion leads to a completely different result than the dynamic heater-like quench initiation with 1D heat diffusion. Most of the resistive voltage is generated by the coil turns that are in the current sharing regime due to increased temperature due to 3D heat diffusion. Fig. 10 shows the temperature of the coil for the two cases once the resistive voltage reached 0.1 V. In the 1D case with the heater, the resistive voltage is generated only by a part of a single turn with a peak temperature of 91 K, which is followed shortly by a thermal runaway. For the 3D case with the defect, the voltage is generated by several turns, which are heated up by the defect, and the peak temperature is 19 K. As shown in Fig. 8, the peak temperature will only slowly increase for tens of seconds before the winding finally enters the thermal runaway regime.

This highlights that it is necessary to carefully consider what a realistic reason for a quench of an HTS coil is. It is still possible that a transient event heats up a small fraction of a turn and makes the detection of this type of quench very challenging. On the other hand, a quasi-static quench caused by a local $I_c$ defect, is likely to be detected in time to prevent a thermal runaway.

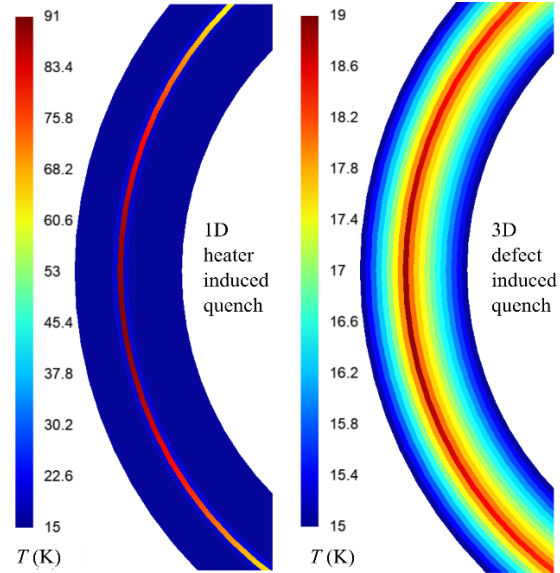

Fig. 10 Temperature distribution at times when the resistive voltage reached the quench detection threshold of 0.1 V for the 1D case with the heater quench initiation (left) and the 3D case with the $I_c$ defect (right).

The impact of coil conduction cooling is likely considerable here, and a different conclusion could be reached when the coil is bath-cooled. This is out of the scope of this paper.

## V. CONCLUSION

The recently developed FiQuS tool was used to study a high-temperature superconductor coated-conductor pancake coil. The concept of minimum defect critical current was introduced, and its dependence on defect length and position within the coil, as well as the effects of 1D and 3D heat diffusion in a conduction-cooled environment, were analyzed. The influence of the coil's magnetic field on these results was also demonstrated. This approach, when applied to a specific coil, could be valuable for decision-making when determining the acceptable critical current defect based on available critical current measurements along the coated conductor length.

Additionally, it was shown that for a conduction-cooled coil with a defect leading to thermal runaway, the resistive voltage exceeds the typical quench detection threshold for several tens of seconds before thermal runaway occurs. This finding differs significantly from the 1D heat diffusion case with the quench initiation using a heater, which is a typical case for determining minimum quench energy. In this case, there is minimal time to validate the signal and activate protection measures before the peak temperature rises to critical levels. Both cases are potentially valid, and careful consideration is required to determine the most realistic quench cause for a given magnet. The consequences of this decision, as well as the challenges in detecting and protecting HTS coils, have been clearly demonstrated.